\documentclass[rmp, sort&compress, numbers]{revtex4} 
\usepackage{graphicx}
\usepackage{rotate}
\usepackage{epsfig}

\usepackage{color}

\usepackage{amssymb,amsfonts,amsmath}

\begin{document}

\title{The Emergence of Consensus: A Primer}

\author{Andrea Baronchelli}
\email{Andrea.Baronchelli.1@city.ac.uk}
\affiliation{City, University of London} 

\begin{abstract}
The origin of population-scale coordination has puzzled philosophers and scientists for centuries. Recently, game theory, evolutionary approaches and complex systems science have provided quantitative insights on the mechanisms of social consensus. However, the literature is vast and widely scattered across fields, making it hard for the single researcher to navigate it. This short review aims to provide a compact overview of the main dimensions over which the debate has unfolded and to discuss some representative examples. It focuses on those situations in which consensus emerges `spontaneously' in absence of centralised institutions and covers topic that include the macroscopic consequences of the different microscopic rules of behavioural contagion, the role of social networks, and the mechanisms that prevent the formation of a consensus or alter it after it has emerged. Special attention is devoted to the recent wave of experiments on the emergence of consensus in social systems.
\end{abstract}

\maketitle

\section{Introduction}

Money, language, dress codes, decorum, notions of fairness all need to be accepted and shared at the group level in order to function. They require social consensus and in exchange they provide individuals with expectations on how others will behave, eventually allowing a society to operate \cite{lewis1969convention,hechter2003theories}. But how does consensus (or `order', `coordination', `agreement') emerge out of an initially disordered situation when there is more than one possible equilibrium? 

This question is key to the Social Sciences and to a wide array of disciplines, ranging from Biology to Physics and from Ethology to Artificial Intelligence. In fact, nature offers countless examples of initially disordered collections of agents that are able to develop shared coordinated behaviours. Flocks of birds frequently change their flight direction \cite{attanasi2014information}, fish schools display spontaneous evasion waves \cite{rosenthal2015revealing}, ferromagnetism is the result of ordering spins \cite{peierls1936ising} and designing decentralised artificial systems is one of the frontiers of Robotics \cite{werfel2014designing}. All these areas have contributed to advance our comprehension of the mechanisms of consensus \cite{skyrms1996evolution,hechter2003theories,castellano2009statistical,ehrlich2005evolution,young2015evolution}.

This interdisciplinary interest has determined the recent explosion in the number of scientific articles investigating the emergence of consensus, with two consequences.
On the one hand, the similarity between explanations proposed in different areas risks to go unnoticed due to different jargons and problem-specific details. On the other hand, communities of researchers exist that largely ignore each other even within apparently confined contexts. For example, two recent and insightful reviews mainly concerned with the problem of consensus such as  ``The Evolution of Norms'' \cite{ehrlich2005evolution} and ``The Evolution of Social Norms'' \cite{young2015evolution} do not share a single bibliographic entry. At the same time, in the last few years it has become clear that understanding the interplay between social consensus and our collective behaviour is crucial to address many of the issues faced by our complex society such as climate change, biodiversity loss and antibiotic resistance \cite{nyborg2016social}.

A systematic review of the literature on the emergence of consensus is out of the scope of the present paper, which offers a very brief introduction to the subject.
My aim is to provide an overview of the most important principles over which the debate on social consensus has unfolded, and to discuss their implications with the aid of few \textit{illustrative examples}. By adopting the language of social conventions, possibly the simplest example of social consensus taken from the Social Sciences (Sec.÷\ref{sec:conventions}), I will start by mapping the landscape of proposed solutions to the problem of consensus (Sec.÷\ref{sec:dimensions}) before focusing on the case of spontaneous emergence in absence of a centralised authority (Sec.÷\ref{sec:spont}). In this context, by considering two simple models, I will discuss how different kinds of behavioural contagion and social networks influence the dynamics of collective agreement (Sec.÷\ref{sec:micromacro}), as well as which mechanisms can either alter (Sec÷\ref{sec:committed}), hinder or prevent consensus (Sec.÷\ref{sec:nocons}). Finally, I will overview recent experiments that provide empirical basis to the study of the emergence of consensus in social systems (Sec.÷\ref{sec:exp}).

\section{A prototypical example: Social Conventions}
\label{sec:conventions}

The word ``spam'' refers to ``disruptive online messages [..] sent as email'' \cite{dictionary2014dictionary}. However, the Internet and - as a consequence - the phenomenon that today we indicate as spam did not exist just a few decades ago. So how did we end up agreeing that those annoying messages are to be called ``spam''? Or, actually, how did we \textit{manage to} agree? 

Naming conventions have attracted the attention of philosophers since the ancient past. Hermogenes, in Plato's Cratylus, asserts that names belong to things ``only because of the rules and usages of those who establish the usage and call it by that name'' \cite{sedley2003plato}, without commenting on \textit{how} a group reaches consensus on a specific name. On the other hand, Adam, the first human in the Bible, establishes new names for the objects around him \cite{jacob2007first}. Far from being curiosities, these two solutions identify a first major divide between different approaches. Consensus can be imposed by an authority or emerge from an interacting multitude.

Conventions govern much of social and economic life. In general, a convention is a pattern of behaviour that is customary, expected and self-enforcing \cite{lewis1969convention,young1996economics}. It is the result of a coordination process where one among various different alternatives is adopted, and they are maintained because a unilateral deviation makes everyone worse off  \cite{young1996economics}. Among the countless situations where consensus plays a crucial role, some of which have been mentioned above, this review adopts the perspective - and language - of social conventions both for its transparency for readers with different backgrounds and for its historical prominence.

\section{Modelling consensus}
\label{sec:dimensions}

The emergence of consensus can be described both as a \textit{cooperative process} in the space of individuals trying to coordinate with each other, and as a \textit{competitive process} in the space of the alternatives individuals can adopt. Different approaches make different hypothesis on the structure of these two spaces. Here we will consider only theories that describe consensus as the result of the interactions between individuals \cite{fields1995elementary,mead1934mind}, but different possibilities exist \cite{o1994marx,sedley2003plato,pinker1995language}.

\subsection*{Space of individuals}

A population can be described in terms of a network whose nodes represent individuals and links identify potential interactions. The coordination between groups of neighbouring nodes is referred to as \textit{local} consensus, while \textit{global} consensus indicates that (most of) the population has reached an agreement. The structure of the social network plays a major role on the dynamics of consensus, in ways that depend on the details of the microscopic individual interactions (see Section \ref{sec:micromacro}). A major distinction between different models of consensus concerns the presence and role of a formal or informal centralised institution, or in general of any actors or mechanisms able to exert a global influence on the system. 
\begin{enumerate}
\item Prominent examples of centralised institutions are:
\begin{itemize}
\item Authority. An authority that has the means to enforce order through violent or non-violent punishment of the violators is the simplest source of social order \cite{hobbes2006leviathan}.
\item Leadership. Leaders need to be identified as such based on some merit \cite{myers1972dimensions}. Potential leaders include `connectors', who have a large social circle, `meavens', who rely on a deep knowledge of a specific topic, and `persuaders', who have exceptional negotiation skills \cite{gladwell2000tipping}.
\item Broadcasting. One-to-many distributors of information can influence consensus both on a specific opinion or by `setting the agenda' on a set of acceptable or urgent problems \cite{mccombs1972agenda,mccombs1997building}.
\item Explicit incentives for collective coordination. A centralised institution makes individuals aware that they will benefit from global consensus, potentially making them more prone to seek coordination also outside of their immediate social circle \cite{kearns2009behavioral,oliver1993formal}.
\item Informational feedback. While no incentives for global coordination exists in this case, individuals are informed about the population-level popularity of the different options \cite{merton1951social}. Conformity and social pressure can then favour final consensus \cite{asch1955opinions}.

\end{itemize}

\item When a centralised institution does not exist, consensus comes either from the interaction between agents or from some pre-defined individual behaviour. Examples of the two cases are: 
\begin{itemize}
\item Spontaneous emergence of consensus. Consensus is said to be ``spontaneous'' when a centralised institution is not present and agreement is produced by self-interested individuals who are not intentionally aiming to global coordination \cite{sugden1989spontaneous}. The dynamics of the process, or `evolutionary' forces \cite{young2015evolution}, select the equilibrium \cite{lewis1969convention,axelrod1986evolutionary,sugden1989spontaneous,young1993evolution,steels1995self,bicchieri2005grammar,baronchelli2006sharp}. Important mechanisms that can foster spontaneous consensus are \cite{bikhchandani1992theory}
\begin{itemize}
\item Communication. For example, earlier participants can explain the benefits of coordination to latecomers \cite{rogers2010diffusion}, or individuals can negotiate some form of local consensus \cite{wittgenstein1958philosophical,garrod1994conversation}.
\item Peer punishment of deviants. When the benefit of (local, at least) consensus are greater than the individual cost of punishing her peers or if the cost of being punished is large enough, then sanctions on deviants are a powerful tool to promote consensus \cite{boyd1992punishment}.
\item Positive payoff externalities. This is the case of self-enforcing norms, such as for example driving on the left or on the right of the road. Once established they persist indefinitely \cite{schelling1960strategy,dybvig1983adoption,farrell1985standardization,katz1985network,arthur1989competing,valverde2015punctuated}.
\item Conformity bias. An inherent tendency to conform to the behaviour of others is a hallmark of human culture \cite{asch1955opinions,jones1984economics,rb2005} and has been observed also among chimpanzees \cite{whiten2005conformity}.
\end{itemize}
\item Quorum sensing. Individuals are capable of assessing the number of peers they interact with and share a pre-defined response once a threshold number of components is detected. Bacteria \cite{miller2001quorum}, ants \cite{mallon2001individual} and honey bees \cite{seeley2004group} are examples of social species that use quorum sensing. 
\end{itemize}

\end{enumerate}

\subsection*{Space of alternatives - Equilibrium selection}

A natural question is which alternative, or equilibrium, will be selected by the population in case of consensus. Three notable answers are:

\begin{itemize}
\item Individuals select a given alternative by logical reflection. They are able to assess the advantages of one equilibrium over the others and act to maximise their benefit. Rational considerations would therefore guide individual choice \cite{harsanyi1988general,von2007theory}.
\item  Individuals select a given alternative based on psychological, even though not rational, factors. Shared biases select the best alternative to be played \cite{schelling1960strategy,lewis1969convention}. 
\item Alternatives are equivalent, and the dynamics of the process where learning individuals interact eventually selects one of the possible equilibria ``by chance'' \cite{skyrms1996evolution,young1996economics,steels1995self}. 
\end{itemize}

Notice that only in the latter case communication, or interaction, between individual is necessary to reach a consensus, as we will see in Sec.~\ref{sec:spont}. In the other two cases, in fact, individuals independently select the same alternative based on internal processes \cite{lewis1969convention}. A further dimension affecting the three scenarios concerns the \textit{basin of attraction} of the different alternatives, i.e. the region of the phase space such that any point (any initial condition) in that region will eventually be iterated into the attractor \cite{ellison2000basins,samuelson1998evolutionary}.

\section{Approaches to the study of spontaneous consensus}
\label{sec:spont}

This and the following sections focus on the case of spontaneous emergence of consensus, where the aim is to understand the macroscopic consequences of microscopic behaviours  \cite{ehrlich2005evolution,castellano2009statistical}. Two main approaches to investigate spontaneous consensus are game theory and the evolutionary - or dynamic - approach.

In coordination games with multiple pure-strategy Nash equilibria, consensus emerges when one equilibrium is selected by all the members of the population. However, it was soon realised that traditional game theory fails to explain how players would know that a Nash equilibrium is to be played and which Nash equilibrium is to be selected when more than one equivalent choices are present \cite{kandori1993learning}. A possible solution is attributing the equilibrium selection at the level of individual decision making \cite{harsanyi1988general}, but this requires strong and unrealistic assumptions on the individual access to, and processing of, information \cite{binmore1987modeling,simon1996sciences}. 

Evolutionary explanations overcome this difficulty by substituting actors' rationality and knowledgeability with the capability of anticipating what others will do, and by specifying how individuals learn from experience and adjust their choices accordingly. At least two main frameworks implement this approach. On the one hand, in Evolutionary Game Theory \cite{smith1973lhe} individuals, who are born with a behavioural strategy, interact and reproduce according to a fitness proportional to the payoff of the game they play. Evolution determines over time the successful strategies, possibly driving the population to an equilibrium. Crucially, the biological framing of genetically encoded strategies and reproduction can be translated in terms of bounded rationality and learning when describing social systems
\cite{foster1990stochastic,young1993evolution,young2015evolution}. On the other hand, agent based modelling aims to understand the global consequences of individual adaptive behaviour relying on the concepts of emergence and self-organisation developed in Statistical Physics. Pioneered by celebrated examples such as Schelling's segregation model \cite{schelling1971dynamic}, Axelrod's work on competition and collaboration \cite{axelrod1997complexity} and Reynold's flocking model \cite{reynolds1987flocks}, agent based models have witnessed an explosion in the last two decades thanks to computational methods and numerical simulations, acquiring a central role in the study of social dynamics. A review of these models is out of the scope of the present paper, which is condemned to leave out many interesting contributions, but the interested reader can found an extensive survey in \cite{castellano2009statistical}.

\section{Microscopic interactions, social networks and the dynamics of consensus}
\label{sec:micromacro}

Multi-agent models define agents that can assume different states, and rules that determine how these states change, typically through interactions. A major distinction concerns the number of times an agent needs to be exposed to another state before adopting it. In simple contagion models, one exposure to a different state may be sufficient \cite{pastor2015epidemic}. In complex contagion models, on the other hand, more exposures are required, typically from more than one source (if interactions reveal the identity of the individuals) \cite{centola2007complex}. The consequences of the adopted kind of contagion can be profound and it is useful to see it in two simple models, chosen purely as illustrative examples.

The Moran process was introduced to study selection in a finite population \cite{moran1958random}. Individuals are characterised by a state variable that can assume one of $M$ values. In each time step two neighbouring individuals are randomly chosen, one for reproduction and one for elimination. The offspring of the first individual will replace the second. Equivalently, it can be said that the second individual will adopt the state of the first one, in a process of \textit{simple contagion}. The same dynamics was introduced a second time under the name of voter model \cite{clifford1973model,holley1975ergodic}. Here, the first individual adopts, or `copies', the state of the second one. The two variants are equivalent on homogeneous topologies but exhibit different behaviours on heterogeneous networks \cite{suchecki2005voter,castellano2005effect,sood2005voter,sood2008voter}.

The naming game addresses the emergence of simple (linguistic) conventions following a scheme devised by Wittgenstein \cite{wittgenstein1958philosophical,steels1995self,baronchelli2006sharp}, which is very similar to the signalling game introduced by Lewis \cite{lewis1969convention} when decisions based on common knowledge are replaced by adaptive behaviour \cite{barr2004establishing}. In the nowadays standard formulation \cite{baronchelli2006sharp}, individuals are characterised by an inventory of names, which is empty at the beginning of the process. In each time step a pair of neighbouring agents is chosen randomly, one to play as hearer and the other a speaker. The speaker randomly selects one of its names, or invents a new name if its inventory is empty. If the hearer's inventory contains such a name, the two individuals update their inventories so as to keep only the word involved in the interaction, otherwise the hearer adds the name to those already stored in its inventory. Thus, at least two interactions are needed for an individual to go from state $A$ to state $B$, a characteristic feature of \textit{complex contagion}.  The number of names can be fixed by endowing agents with a name at the beginning of the game. 

\begin{figure}[t]
\includegraphics*[width=0.6\textwidth]{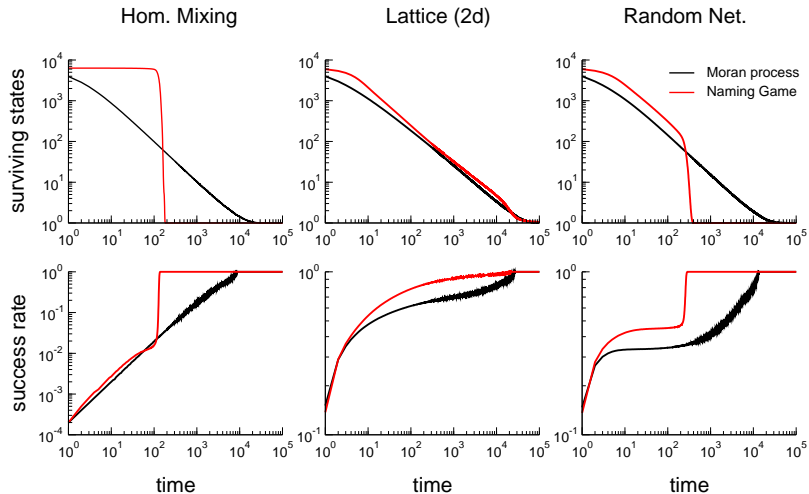}
\caption{\textbf{Different paths to spontaneous consensus.} Top: surviving states for the Moran process (simple contagion) and naming game (complex contagion) on different topologies. Bottom: Success rate, defined as the probability of observing an interaction involving two identical individuals in the Moran process or a successful interaction in the naming game (similar alternative observables exist for the two models, the qualitative description is not affected by the particular choice). In homogeneously mixing populations the Moran process evolves through a progressive elimination of different states, while the naming game exhibits a sharp transition to order (symmetry breaking). The dynamics of the two models appear more similar on lattices, although profound differences exist (see Fig. \ref{fig2}). On complex networks, on the other hand, after an initial phase in which the two models appear similar, the naming game exhibits a transition to order similar to the one observed on Homogeneously mixing populations. Population size of $N=10,000$ individuals prepared initially in $M=N$ different states. Lattice and random network have coordination number $k=4$ for all the nodes.}
\label{fig1}
\end{figure}

In finite size populations, consensus emerges both in the Moran process and naming game and in both cases once it is reached it will persist indefinitely. However, the mechanisms controlling how the population `selects' the alternative to agree upon are qualitatively different in the two models. To see this, it is convenient considering different interactions topologies separately. The Appendix contains a glossary of network terms. 

Before proceeding, it is worth noting that beyond the number of exposures necessary for an agent to change state other factors play an important role. According to the theory of social impact \cite{latane1981psychology}, for example, the impact of a group on an individual is proportional to the ``strength'' of the members of the group (how credible or persuasive they are), their ``immediacy'' (a decreasing function of their social ``distance'' from the individual), and their number, $N$. Various multi-agent models explored \cite{nowak1990private, lewenstein1992statistical,holyst2000phase} or took inspiration from \cite{schweitzer2000modelling} this approach but unfortunately we are forced to limit our analysis to the above mentioned examples due to space limitations.

\subsubsection{Homogeneously mixing populations}
In the Moran process, interactions are symmetrical. Thus, if only two states are available, when an agent in state state $0$ and an agent in state $1$ interact, the outcome is either two agents in $0$ or two agents in $1$ with the same probability ($p=1/2$). Thus, a chain of interactions favouring one state (i.e., a large \textit{fluctuation}) is needed in order for that state to prevail. When more states are available, consensus is reached through a progressive elimination of alternatives (Fig. \ref{fig1}). In particular, the probability that consensus is reached on state $1$ when there are $i$ many $A$ individuals in the population is simply $i/N$ \cite{nowak2006evolutionary}. Hence, at any time before consensus there is a probability $(N-i)/N$ that $0$ will dominate. The expected number of interactions per individual needed to reach consensus is proportional to the population size $N$ \cite{cox1989coalescing}. 

In the binary naming game agents are initially assigned with one of two names (e.g., $A$ or $B$) and can find themselves in one of the three states identified by an inventory that contains only name $A$, only name $B$ or both $A$ and $B$ \cite{baronchelli2007nonequilibrium}. 
A contact between $A$ ($B$) and $AB$ will increase the population of $A$ ($B$) with probability $p=3/4$. Thus, the larger the fraction $n_A$ ($n_B$) of individuals who only know name $A$ ($B$) the more that fraction will increase. Mathematically, the difference between $n_A$ and $n_B$ (notice that $n_{AB}=1-n_{A}-n_{B}$) evolves according to $\frac{d (n_{A}-n_{B})}{dt} \propto (n_{A}-n_{B})$, meaning that the larger faction will always impose its consensus in large populations \cite{baronchelli2007nonequilibrium}.  When the number of states is not restricted, the dynamics is characterised by an initial phase of competition between names, followed by a winner-take-all regime in which the most popular convention progressively eliminates all the competitors \cite{baronchelli2006sharp}, in a process known as 
\textit{symmetry breaking} (Fig. \ref{fig1}).
The time needed to reach consensus is faster than in the Moran process, and proportional to $\log N$ and $\sqrt{N}$ interactions for the binary and the unrestricted models, respectively \cite{baronchelli2006sharp,baronchelli2008depth}.

\begin{figure}
\includegraphics*[width=0.9\textwidth]{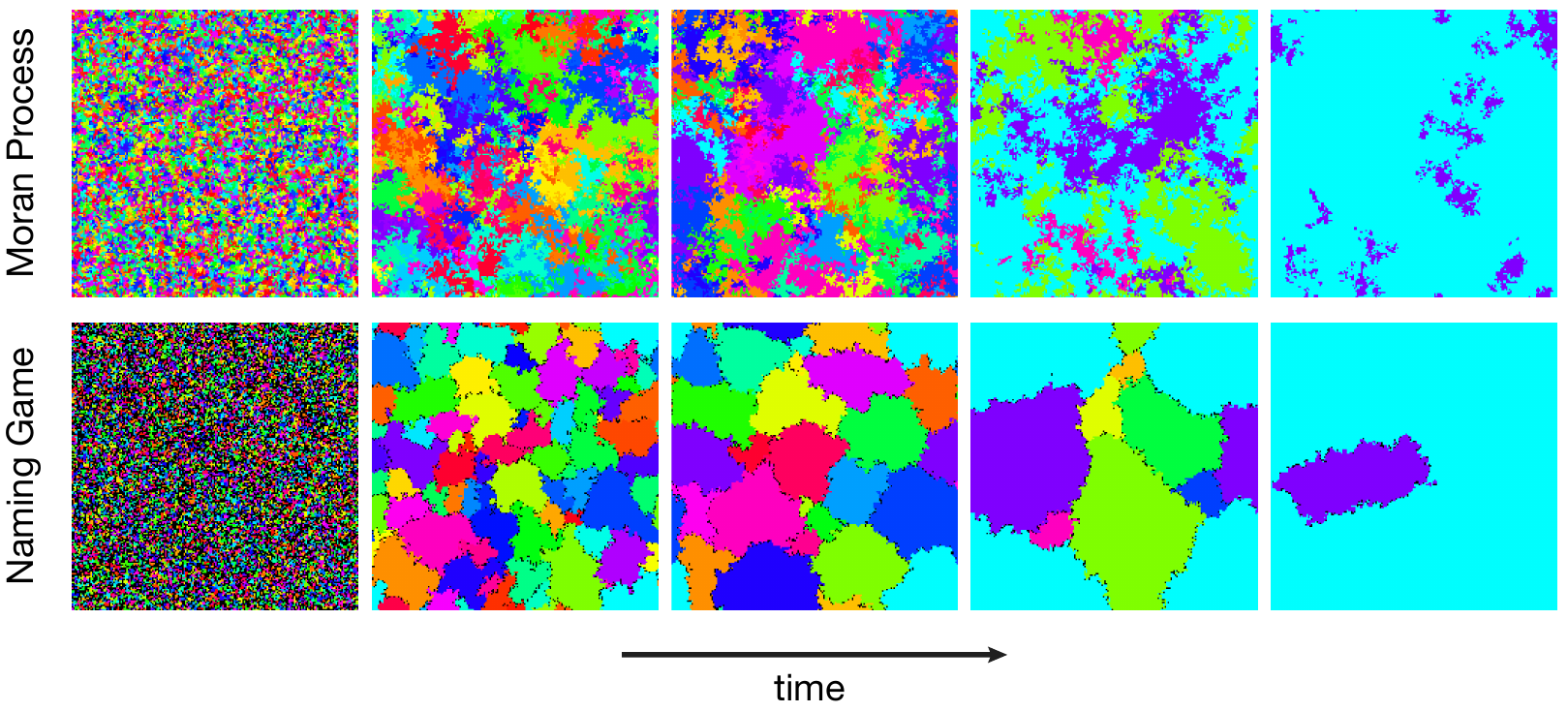}
 \caption{\textbf{From local to global consensus in spatial networks.} Snapshots of the temporal evolution of the Moran process (simple contagion, top panels) and naming game (complex contagion, bottom panels) on a 2-dimensional lattice with coordination number $4$ and periodic boundary conditions. While compact clusters of agreeing agents form in the naming game, in the Moran process regions of the same colour are difficult to identify and often broken in more pieces. Population of $N=40,000$ agents, initial condition with $M=N$ different states (i.e, each agent starts in a different state). Colours correspond to different states, with the exception of the left panels where for visualisation purposes it is possible that different states are rendered in the same colour. Black points in the naming game correspond to agents with more than one name in their inventory.} 
 \label{fig2}
 \end{figure}

\subsubsection{Spatial networks}

On two-dimensional regular lattices the time required to reach consensus  is $t_{consesus} \sim \ln N$ for the Moran process \cite{cox1989coalescing} and $t_{consesus} \sim N$ for the naming game \cite{baronchelli2006topology}. While Figure \ref{fig1} might suggest that the dynamics of the two models is similar on lattices, Figure \ref{fig2} shows that important differences exist.
In the naming game, local consensus between neighbouring individuals emerges rapidly but different regions reach a consensus on different conventions. Clusters of \textit{local} consensus stay compact and the dynamics proceeds trough cluster-cluster competition at the frontier between different regions. In the Moran process, on the other hand, simple contagion prevents the formation of such compact clusters, and the path to global consensus is dominated by fluctuations as in homogeneously mixing populations. 

It is important to note that, beyond these two models, other scenarios exist and opposite results can be found. For example, in the context of the coordination game with bounded rationality of \cite{kandori1993learning} convergence to the risk-dominant strategy is slow on fully connected graphs, where initial conditions play a predominant role, while evolutionary forces determine the outcome when players interact with small sets of neighbours in clustered networks \cite{ellison1993learning}.

\subsubsection{Complex networks}

Most networks observed in nature are characterised by the small-world property \cite{travers1967small,watts1998collective}, describing the fact that the average distance between any pair of nodes grows as the logarithm of the system size, and a broad distribution of node connectivity $k$ \cite{barabasi1999emergence}, often compatible with a scale free behaviour $P(k) \sim k^{-\gamma}$ with $2<\gamma<3$ \cite{barabasi1999emergence,newman2003structure,caldarelli2007scale}. On such topologies, both the Moran process and the naming game recover the behaviour and scaling exponents observed in homogeneously mixing populations \cite{suchecki2005voter,castellano2005effect,dall2006nonequilibrium}. In the naming game, after an initial phase of local agreement, the small world property favours the spreading of conventions between different regions thus preventing the formation of regional clusters \cite{dall2006agreement}. 

However, on scale-free networks Moran and voter model behave differently, the presence of hubs slowing down consensus in the Moran process and favouring it for the voter model \cite{suchecki2005voter,castellano2005effect,sood2005voter,sood2008voter,vazquez2008analytical}. In general, in any model describing pairwise interactions, the topology and role of the agents become entangled on heterogeneous networks. The first individual is selected according to the degree distribution $P(k)$ while the second individual, being selected among the neighbours of the first one, is sampled from a different distribution, which in the case of uncorrelated networks is $Q(k) \sim kP(k)$ \cite{newmannetworks}. 

Of course, further possibilities exist and predictions of game-theoretic models may be antithetic to the ones described above. For example, innovations spread quickly in locally connected networks and geographical networks, while hubs are an obstacle to the spreading of a risk-dominant strategy in a model where the payoff of each alternative increases with the number of neighbours who are adopting the same choice  \cite{montanari2010spread}.

\section{Fragile Consensus and Committed Minorities}
\label{sec:committed}
The large majority of models describe consensus as an absorbing state: once reached, it will persist indefinitely \cite{castellano2009statistical,galam2008sociophysics}. However, social consensus is often fragile. Apparently small shocks or weak forces can result in global shifts of behaviour, causing consensus to move from one equilibrium to a different one. Cohabitation of unmarried couples, same-sex relationships, and social attitudes towards legal and illegal drugs have changed over the course of the last decades \cite{kuran1989sparks,bikhchandani1992theory}. Interestingly, often the transition from one equilibrium to the new one is swift, and the reshaping of consensus can be described in terms of physical concepts such as a phase transition \cite{haken1975cooperative,gladwell2006tipping} or a collective swing due to spontaneous fluctuations \cite{cavagna2016non}.

An important question is whether a small fraction of committed actors can push the majority of the population towards a different equilibrium. Various social phenomena, from revolutions \cite{gladwell2010small} to the constant renewal of current day slang \cite{lighter1994random} and to fashions and fads \cite{bikhchandani1992theory} are in fact attributed to the activity of initially small groups. The two models we have examined above have been extensively studied in this context, following pioneering insights from different approaches \cite{galam2007role}.  

In the case of the Moran process, even a minority of non-committed individuals has always a chance to sway the majority opinion.  It is remarkable, however, that a single committed agent is able to lead the whole population towards the state it chooses in spatial lattices, while it is unable to do so in higher dimensions \cite{mobilia2003does,mobilia2003majority}. In the binary naming game, on the other hand, we have seen that the majority opinion will always be imposed at the population level. However, it can be shown that a minority of individuals committed on name $B$ will be able to flip the consensus reached on $A$ provided that its size exceeds a threshold of around $10\%$ of the individuals \cite{xie2011social,marvel2012encouraging,mistry2015committed}. A similar threshold has been observed also in radically different models \cite{halu2013connect}, while a more heterogeneous distribution of individual commitments yields to minority thresholds in the range between $10\%$ and $40\%$ in the context of the naming game \cite{niu2017impact}.

\section{Obstacles to spontaneous consensus and coexistence of different states}
\label{sec:nocons}
Most formal models of social influence seem to imply that consensus is unavoidable \cite{shao2009dynamic,mas2010individualization,abelson1964mathematical,li2013non}. However, disagreement characterises many aspects of our society. A natural question is therefore what factors can hinder the process of consensus in models that would otherwise lead to it. 

One natural answer is topology. Networks characterised by a strong community structure can enormously slow down, or even prevent, consensus in models of complex contagion \cite{dall2006nonequilibrium,lambiotte2007coexistence,lu2009naming}. The same mechanisms yielding compact clusters of agreeing individuals in spatial networks guarantee the cohesiveness of a topological community. Furthermore, simple modifications of the microscopic rules, such as a nonlinear dependence of the transition rates on the states of neighbouring nodes in the voter model \cite{schweitzer2009nonlinear} or an irresolute attitude of the agents in the naming game \cite{baronchelli2007nonequilibrium}, may guarantee the coexistence of different states even on lattices or fully connected graphs.

A different mechanism is proposed by the well-known Axelrod model of dissemination of culture \cite{axelrod1997dissemination}, defined 
as a set of individual attributes that are subject to social influence. Given that individuals have a tendency to interact more with others who share their opinion (homophily) and that interactions between individuals tend to increase their similarity (social influence), where do cultural differences come from? The answer has to be sought in the mechanisms of `bounded confidence' according to which only individuals that are already sufficiently similar interact \cite{deffuant2000mixing}. In the model, individuals are characterised by $F$ cultural features that can assume $q$ traits. At each time step, two individuals are randomly selected and interact with a probability proportional to the number of features for which they share the same trait. The result of an interaction is that the two individuals will increase their similarity by aligning one feature for which traits are different. If the number of possible traits, $q$ is small the process will end up in a state of consensus where all individuals share the same trait for the same feature, but a threshold value exists such that for $q>q_c$ consensus will not be reached \cite{castellano2000nonequilibrium,klemm2003nonequilibrium}. Furthermore, it has been shown that the interplay between local interactions and the homogenising effect of a centralised ordering effort produces non trivial results and may increase the disorder the system \cite{gonzalez2005nonequilibrium,gonzalez2006local}.

Interestingly, it has recently been shown that topology and homophily interact in online social networks, where users have the possibility to control who to connect to (see \cite{centola2007homophily} for the modelling of this feature in the context of the Axelrod model). Here, tightly connected and relatively isolated communities emerge spontaneously, maintaining and promoting group polarisation. These `echo-chambers' hinder consensus not only at the level of social conventions and norms, but also on the recognition of e.g. scientific evidence \cite{del2016echo} with consequences on public debate \cite{del2016spreading}. Theoretical approaches including dynamical network modelling along with homophily and social influence confirm this picture \cite{starnini2016emergence}. An open question, whose urgency has been stressed also by the World Economic Forum, is what can be done to favour a less polarised debate in our society (see also Table \ref{t:questions}) \cite{howell2017global,zollo2015debunking}.

\section{Empirical studies}
\label{sec:exp}
Insights on the emergence of consensus often come from studies designed with a different focus. This section covers some examples representative of different approaches, motivations and implementation schemes.

Language is a natural environment for the study the spontaneous emergence of conventions. While the space of alternatives is naturally rich, however, experiments in this context have often aimed to explore coordination on higher linguistic features (e.g., the emergence of compositionality) involving small population sizes. Galantucci investigated the emergence of a communication code in a simple coordination game  \cite{galantucci2005experimental}. Pairs of physically separated individuals had to coordinate on where to go in order to meet in a simple set of communicating room they see on a screen. Communication was mediated by a system that does not allow users to write (a sliding trackpad). The author found that a communication system emerges, signs may originate from different mappings (movement, position etc), systems develop parsimoniously (new signs are related to already established signs) and final signs were well distinct. Related yet different experiments showed that an unstable environment may facilitate the emergence of sophisticated forms of coordination, such as a compositional code, when pairs of individuals communicate \cite{selten2007emergence}. Garrod and Doherty analysed the role of a community, as opposed to just 2 communicating individuals, where individuals - interacting in pairs - had to describe their changing position in a maze \cite{garrod1994conversation} . The presence of more users (up to $N=10$, in the experiments) slowed down the initial agreement, but resulted in a more stable consensus, i.e. in a final state with more successful interactions based on more stable codes, in agreement with Lewis view of conventions as solutions to \textit{collective} coordination problems \cite{lewis1969convention,lewis1975languages,garrod1987saying}.

The spontaneous emergence of consensus was explicitly addressed in \cite{centola2015spontaneous} through a coordination game played by group sizes of up to $N=96$ individuals. In a given round of the game, two network ÒneighboursÓ were chosen at random to play with one another. Both players simultaneously assigned names to a human face. If the players coordinated on a name, they were rewarded with a successful payment, otherwise they were penalised. After a single round, the participants could see only the choices that they and their partner had made. They were then randomly assigned to play with a new neighbour in their social network, and a new round would begin. The object that participants were trying to name was the same for the entire duration of the game, and for all members of the game. The experiments showed that global consensus emerges in homogeneously mixing populations, while different clusters of local consensus appear in spatial networks in agreement with the predictions of the naming game model \cite{baronchelli2006sharp}.

Kearns et al. explored the problem of consensus in presence of an explicit incentive for collective agreement \cite{kearns2009behavioral}. A population of $N=36$ individuals was arranged on networks with different topologies to play a networked version of the classic `Battle of the Sexes' game. Individuals were in one of two possible states, labelled `red' and `blue', and their payoff in the game depended on which state will eventually be adopted by the whole population. Each individual knew the state of their neighbours in the network and could change colour at each time step. Results showed that when incentives were randomly distributed in the population, so that $50\%$ of the individuals prefer blue and the other $50\%$ prefer red, consensus is reached in only $57\%$ of trials. When, on the other hand a certain payoff was assigned to a minority of individuals occupying well connected nodes in a heterogeneous network consensus is much more likely ($89\%$ of trials). Judd et al. adopted a similar setting, where global agreement was the explicit goal and individuals characterised by a simple colour variable have access to the state of their neighbours, in the experiments reported in \cite{judd2010behavioral}. Starting from a network characterised by a strong community structure ($6$ communities for a a population of $N=36$ individuals), these experiments confirmed that `long-distance' connections, i.e. the small-world property, promote consensus. 

\begin{table}
\begin{tabular}{p{15cm}}
\hline
  \begin{itemize}
\item Can behavioural change be engineered? Can we foster social consensus on beneficial behavioural norms, such as practices of environmental sustainability or social inclusion? Conversely, how can negative yet widespread norms - from bullying to corruption - be eradicated?
\item How can we contrast the formation of online self-organising communities, or `echo-chambers'? How can the connectivity of a social network be increased? How robust are these echo-chambers? How do overlapping echo-chambers interact?
\item How can committed minorities be put to use to induce social change? Can their role be tested in the lab? Are the properties of the network structure in social interactions a key factor for the effectiveness of committed minorities?
\item How are online social networks changing the mechanisms of social consensus? What is the interaction between online and offline paths to consensus? What is the role of centralised and decentralised mechanisms of information production on the formation of social consensus? How does consensus emerge on new ways to attribute and store values such as cryptocurrencies? 
\end{itemize}
\\
\hline
\end{tabular}
\caption{Outstanding questions.} 
\label{t:questions}
\end{table}

Empirical investigations of the spreading of behaviour have provided important insights on the existence and nature of complex contagions, which as we have seen is a crucial ingredient of many models for the emergence of consensus. Microscopic complex contagion has been studied in the laboratory \cite{centola2010spread} and in offline \cite{christakis2007spread} and online \cite{hodas2013simple,aral2017} social networks also in relation with its interplay with the topology of the network.  Other experiments have started to unveil previously neglected aspect of the coordination process. For example, the structure of incentives has  been investigated, showing that higher stakes increase the pressure to establish and adhere to shared expectations that persist across rounds \cite{hawkins2016formation}. 

Finally, outside of the laboratory, conventions have been investigated for example using Twitter. Focusing on the adoption (i.e., first use) of markers for retweet or tweet quoting Kooti et al. found that, despite many alternatives were proposed eventually the conventions of `RT' and `via' became dominant. Interestingly, successful conventions were initially proposed and adopted by active and well connected users at the core of the Twitter community, showing that status, influence and connectedness play an important role, changing the ideal condition of interacting \textit{peers} \cite{kooti2012emergence}. Interestingly, a similar role of earliest users in determining the normative consensus has been found also in Wikipedia \cite{heaberlin2016evolution}.

\section{Concluding remarks and outlook}

This overview has necessarily been (very) selective, but it allows us to draw encouraging conclusions. Major advancements occurred in the past years have shed new light on the process of consensus formation. Theoretical milestones in game theory and complexity science have benefited by the steady increase of computational power and the consequent investigation of a large number of models for the study of consensus formation. Different hypotheses have been tested and the micro-macro connection is now much clearer in many situations, although important questions remain open (see Table \ref{Outstanding questions.}). Very recently, finally, empirical approaches, the analysis of human activity on social media and the use of wearable sensors have started to shed light on the mechanisms at play in our society. It is likely that further insights will be produced by the synergy of these three approaches in the next few years.

\section*{Appendix}
\label{Appendix}
This section provides a short definition of some of the terms used in the main text. In some cases, a broader definition exists but only the one useful to an easier reading of the present paper is provided.
\begin{description}
\item[Basin of Attraction:] A region of the phase space of a dynamical system such that initial conditions chosen in that region dynamically evolve to a particular attractor.
\item[Contagion:] Transmission of a disease, idea or behaviour from a person to another by close contact.
\begin{description}
\item[- Simple contagion:] Process in which successive exposures to a pathogen or behaviour are independent and characterised by the same probability $p$ of infection.
\item[- Complex contagion:] Process in which the probability of infection (i.e., for example, adoption of a behaviour) depends on the number of exposures in a complex, non-linear, way.
\end{description}
\item[Nash Equilibrium:] Stable state of a system of interacting individuals in which no player can benefit by changing strategies while the other players keep theirs unchanged.
\item[Self-organisation:] The capability of a system to acquire a functional, spatial or temporal structure without specific interference from the outside \cite{haken}. Sometimes identified with `spontaneous' order in the Social Sciences.
\item[(Spontaneous) Symmetry breaking:] Process of symmetry reduction in a system evolving according to symmetric laws. Arbitrarily small fluctuations drive the system out of the initially symmetrical state and into a final asymmetrical state.
\end{description}

Topology-related terms:
\begin{description}
\item[Homogeneously mixing population:] Population in which agents occupy the vertices of a complete, or fully connected, graph. 
\item[Lattice:]  An arrangement in space of isolated points (lattice points) in a regular pattern. In 2 dimensions, the word `lattice' is typically used to refer to a regular grid in which each point is connected to $4$ neighbouring points.
\item[Network:] A collection of points, called nodes, joined by
lines, referred as links. Vertices represent the elementary
components of a system, for example the individuals in a population, whereas links stand for the possible interactions  between pairs of components (see also \cite{albert2002statistical,boccaletti2006complex,caldarelli2007scale,newmannetworks} for more details on the quantities detailed below).
\begin{description}
\item[- Community:] Although many definitions exist, a community can be generally defined as a set of nodes which are more tightly connected with one another than with other nodes in the network \cite{fortunato2010community}.
\item[- Degree of a node:] The degree $k_i$ of a node $i$ is defined as the number of other nodes to which it is connected, i.e. to the number of its `neighbours'.
\item[- Degree distribution:] The probability $P(k)$ that a randomly chosen vertex has degree $k$.
\item[- Heterogeneous, or `Scale Free', Networks:] Networks with a heavy-tailed degree distribution
that can often be approximated by a power-law, $P(k) \sim k^{-\gamma}$, with $\gamma$ typically between $2$ and $3$. The presence of extremely well connected nodes, or `hubs', is responsible for many of the interesting properties of real-world networks.
\item[- Homogeneous Networks:] Networks with a well-peaked and exponentially decaying degree-distribution, where the variation in connectivity among nodes is limited and hubs are absent.
\item[- Shortest path length,] or distance, between
vertices $i$ and $j$ is the length (in number of edges) of the shortest path joining $i$
and $j$.
\item[- Small-world property:] A property shown by many real networks that
exhibit a small value of the average shortest path length, increasing with
network size logarithmically or slower. This property is in stark contrast to the
larger diameter of regular lattices, which grows algebraically with lattice size.
\end{description}
\end{description}


\clearpage
\begin{thebibliography}{100}

\bibitem{lewis1969convention}
David Lewis.
\newblock {\em Convention: A philosophical study}.
\newblock Blackwell, 1969.

\bibitem{hechter2003theories}
Michael Hechter and Christine Horne.
\newblock {\em Theories of social order: a reader}.
\newblock Stanford University Press, 2003.

\bibitem{attanasi2014information}
Alessandro Attanasi, Andrea Cavagna, Lorenzo Del~Castello, Irene Giardina,
  Tomas~S Grigera, Asja Jeli{\'c}, Stefania Melillo, Leonardo Parisi, Oliver
  Pohl, Edward Shen, et~al.
\newblock Information transfer and behavioural inertia in starling flocks.
\newblock {\em Nature Physics}, 10(9):691--696, 2014.

\bibitem{rosenthal2015revealing}
Sara~Brin Rosenthal, Colin~R Twomey, Andrew~T Hartnett, Hai~Shan Wu, and Iain~D
  Couzin.
\newblock Revealing the hidden networks of interaction in mobile animal groups
  allows prediction of complex behavioral contagion.
\newblock {\em Proceedings of the National Academy of Sciences},
  112(15):4690--4695, 2015.

\bibitem{peierls1936ising}
Rudolf Peierls.
\newblock On ising's model of ferromagnetism.
\newblock In {\em Mathematical Proceedings of the Cambridge Philosophical
  Society}, volume~32, pages 477--481. Cambridge Univ Press, 1936.

\bibitem{werfel2014designing}
Justin Werfel, Kirstin Petersen, and Radhika Nagpal.
\newblock Designing collective behavior in a termite-inspired robot
  construction team.
\newblock {\em Science}, 343(6172):754--758, 2014.

\bibitem{skyrms1996evolution}
Brian Skyrms.
\newblock {\em Evolution of the social contract}.
\newblock Cambridge University Press, 1996.

\bibitem{castellano2009statistical}
Claudio Castellano, Santo Fortunato, and Vittorio Loreto.
\newblock Statistical physics of social dynamics.
\newblock {\em Reviews of modern physics}, 81(2):591, 2009.

\bibitem{ehrlich2005evolution}
Paul~R Ehrlich and Simon~A Levin.
\newblock The evolution of norms.
\newblock {\em PLoS Biol}, 3(6):e194, 2005.

\bibitem{young2015evolution}
H~Peyton Young.
\newblock The evolution of social norms.
\newblock {\em Annual Reviews of Economics}, 7(1):359--387, 2015.

\bibitem{nyborg2016social}
Karine Nyborg, John~M Anderies, Astrid Dannenberg, Therese Lindahl, Caroline
  Schill, Maja Schl{\"u}ter, W~Neil Adger, Kenneth~J Arrow, Scott Barrett,
  Stephen Carpenter, et~al.
\newblock Social norms as solutions.
\newblock {\em Science}, 354(6308):42--43, 2016.

\bibitem{dictionary2014dictionary}
Collins~English Dictionary.
\newblock Dictionary. com.
\newblock {\em Retrieved April}, 15:2017, 2017.

\bibitem{sedley2003plato}
David Sedley.
\newblock {\em Plato's Cratylus}.
\newblock Cambridge University Press, 2003.

\bibitem{jacob2007first}
Benno Jacob, Ernst Jacob, and Walter Jacob.
\newblock {\em The First Book of the Bible, Genesis}.
\newblock KTAV Publishing House, Inc., 2007.

\bibitem{young1996economics}
H~Peyton Young.
\newblock The economics of convention.
\newblock {\em The Journal of Economic Perspectives}, 10(2):105--122, 1996.

\bibitem{fields1995elementary}
Karen Fields and KE~Fields.
\newblock The elementary forms of religious life, 1995.

\bibitem{mead1934mind}
George~Herbert Mead.
\newblock {\em Mind, self and society}, volume 111.
\newblock Chicago University of Chicago Press., 1934.

\bibitem{o1994marx}
Joseph OÕMalley and Richard~A Davis.
\newblock Marx: Early political writings.
\newblock {\em Marx: Early Political Writings}, 1994.

\bibitem{pinker1995language}
Steven Pinker.
\newblock {\em The language instinct: The new science of language and mind},
  volume 7529.
\newblock Penguin UK, 1995.

\bibitem{hobbes2006leviathan}
Thomas Hobbes.
\newblock {\em Leviathan}.
\newblock A\&C Black, 2006.

\bibitem{myers1972dimensions}
James~H Myers and Thomas~S Robertson.
\newblock Dimensions of opinion leadership.
\newblock {\em Journal of marketing research}, pages 41--46, 1972.

\bibitem{gladwell2000tipping}
Malcolm Gladwell.
\newblock {\em The tipping point: How little things can make a big difference}.
\newblock Little, Brown, 2000.

\bibitem{mccombs1972agenda}
Maxwell~E McCombs and Donald~L Shaw.
\newblock The agenda-setting function of mass media.
\newblock {\em Public opinion quarterly}, 36(2):176--187, 1972.

\bibitem{mccombs1997building}
Maxwell McCombs.
\newblock Building consensus: The news media's agenda-setting roles.
\newblock {\em Political Communication}, 14(4):433--443, 1997.

\bibitem{kearns2009behavioral}
Michael Kearns, Stephen Judd, Jinsong Tan, and Jennifer Wortman.
\newblock Behavioral experiments on biased voting in networks.
\newblock {\em Proceedings of the National Academy of Sciences},
  106(5):1347--1352, 2009.

\bibitem{oliver1993formal}
Pamela~E Oliver.
\newblock Formal models of collective action.
\newblock {\em Annual Review of Sociology}, pages 271--300, 1993.

\bibitem{merton1951social}
Robert~K Merton.
\newblock Social theory and social structure: Toward the codification of theory
  and research.
\newblock 1951.

\bibitem{asch1955opinions}
Solomon~E Asch.
\newblock Opinions and social pressure.
\newblock {\em Readings about the social animal}, 193:17--26, 1955.

\bibitem{sugden1989spontaneous}
Robert Sugden.
\newblock Spontaneous order.
\newblock {\em The Journal of Economic Perspectives}, 3(4):85--97, 1989.

\bibitem{axelrod1986evolutionary}
Robert Axelrod.
\newblock An evolutionary approach to norms.
\newblock {\em American political science review}, 80(04):1095--1111, 1986.

\bibitem{young1993evolution}
H~Peyton Young.
\newblock The evolution of conventions.
\newblock {\em Econometrica: Journal of the Econometric Society}, pages 57--84,
  1993.

\bibitem{steels1995self}
Luc Steels.
\newblock A self-organizing spatial vocabulary.
\newblock {\em Artificial life}, 2(3):319--332, 1995.

\bibitem{bicchieri2005grammar}
Cristina Bicchieri.
\newblock {\em The grammar of society: The nature and dynamics of social
  norms}.
\newblock Cambridge University Press, 2005.

\bibitem{baronchelli2006sharp}
Andrea Baronchelli, Maddalena Felici, Vittorio Loreto, Emanuele Caglioti, and
  Luc ~.
\newblock Sharp transition towards shared vocabularies in multi-agent systems.
\newblock {\em Journal of Statistical Mechanics: Theory and Experiment},
  2006(06):P06014, 2006.

\bibitem{bikhchandani1992theory}
Sushil Bikhchandani, David Hirshleifer, and Ivo Welch.
\newblock A theory of fads, fashion, custom, and cultural change as
  informational cascades.
\newblock {\em Journal of political Economy}, 100(5):992--1026, 1992.

\bibitem{rogers2010diffusion}
Everett~M Rogers.
\newblock {\em Diffusion of innovations}.
\newblock Simon and Schuster, 2010.

\bibitem{wittgenstein1958philosophical}
Ludwig Wittgenstein.
\newblock Philosophical investigations, trans. gem anscombe.
\newblock {\em Blackwell, Oxford}, 1958.

\bibitem{garrod1994conversation}
Simon Garrod and Gwyneth Doherty.
\newblock Conversation, co-ordination and convention: An empirical
  investigation of how groups establish linguistic conventions.
\newblock {\em Cognition}, 53(3):181--215, 1994.

\bibitem{boyd1992punishment}
Robert Boyd and Peter~J Richerson.
\newblock Punishment allows the evolution of cooperation (or anything else) in
  sizable groups.
\newblock {\em Ethology and sociobiology}, 13(3):171--195, 1992.

\bibitem{schelling1960strategy}
Thomas~C Schelling.
\newblock {\em The strategy of conflict}.
\newblock Harvard University Press, 1960.

\bibitem{dybvig1983adoption}
Philip~H Dybvig and Chester~S Spatt.
\newblock Adoption externalities as public goods.
\newblock {\em Journal of Public Economics}, 20(2):231--247, 1983.

\bibitem{farrell1985standardization}
Joseph Farrell and Garth Saloner.
\newblock Standardization, compatibility, and innovation.
\newblock {\em The RAND Journal of Economics}, pages 70--83, 1985.

\bibitem{katz1985network}
Michael~L Katz and Carl Shapiro.
\newblock Network externalities, competition, and compatibility.
\newblock {\em The American economic review}, 75(3):424--440, 1985.

\bibitem{arthur1989competing}
W~Brian Arthur.
\newblock Competing technologies, increasing returns, and lock-in by historical
  events.
\newblock {\em The economic journal}, 99(394):116--131, 1989.

\bibitem{valverde2015punctuated}
Sergi Valverde and Ricard~V Sol{\'e}.
\newblock Punctuated equilibrium in the large-scale evolution of programming
  languages.
\newblock {\em Journal of The Royal Society Interface}, 12(107):20150249, 2015.

\bibitem{jones1984economics}
Stephen~RG Jones.
\newblock {\em The economics of conformism}.
\newblock Blackwell, 1984.

\bibitem{rb2005}
Peter~J. Richerson and Robert Boyd.
\newblock {\em Not by Genes Alone: How Culture Transformed Human Evolution}.
\newblock Chicago Univ. press, 2005.

\bibitem{whiten2005conformity}
Andrew Whiten, Victoria Horner, and Frans~BM De~Waal.
\newblock Conformity to cultural norms of tool use in chimpanzees.
\newblock {\em Nature}, 437(7059):737--740, 2005.

\bibitem{miller2001quorum}
Melissa~B Miller and Bonnie~L Bassler.
\newblock Quorum sensing in bacteria.
\newblock {\em Annual Reviews in Microbiology}, 55(1):165--199, 2001.

\bibitem{mallon2001individual}
E~Mallon, Stephen Pratt, and N~Franks.
\newblock Individual and collective decision-making during nest site selection
  by the ant leptothorax albipennis.
\newblock {\em Behavioral Ecology and Sociobiology}, 50(4):352--359, 2001.

\bibitem{seeley2004group}
Thomas Seeley and P~Kirk Visscher.
\newblock Group decision making in nest-site selection by honey bees.
\newblock {\em Apidologie}, 35(2):101--116, 2004.

\bibitem{harsanyi1988general}
John~C Harsanyi, Reinhard Selten, et~al.
\newblock A general theory of equilibrium selection in games.
\newblock {\em MIT Press Books}, 1, 1988.

\bibitem{von2007theory}
John Von~Neumann and Oskar Morgenstern.
\newblock {\em Theory of games and economic behavior (2nd Ed)}.
\newblock Princeton University Press, 1947.

\bibitem{ellison2000basins}
Glenn Ellison.
\newblock Basins of attraction, long-run stochastic stability, and the speed of
  step-by-step evolution.
\newblock {\em The Review of Economic Studies}, 67(1):17--45, 2000.

\bibitem{samuelson1998evolutionary}
Larry Samuelson.
\newblock {\em Evolutionary games and equilibrium selection}, volume~1.
\newblock MIT press, 1998.

\bibitem{kandori1993learning}
Michihiro Kandori, George~J Mailath, and Rafael Rob.
\newblock Learning, mutation, and long run equilibria in games.
\newblock {\em Econometrica: Journal of the Econometric Society}, pages 29--56,
  1993.

\bibitem{binmore1987modeling}
Ken Binmore.
\newblock Modeling rational players: Part i.
\newblock {\em Economics and philosophy}, 3(02):179--214, 1987.

\bibitem{simon1996sciences}
Herbert~A Simon.
\newblock {\em The sciences of the artificial (3rd Edition)}.
\newblock MIT press, 1996.

\bibitem{smith1973lhe}
J~Maynard Smith and GR~Price.
\newblock Lhe logic of animal conflict.
\newblock {\em Nature}, 246:15, 1973.

\bibitem{foster1990stochastic}
Dean Foster and Peyton Young.
\newblock Stochastic evolutionary game dynamics?
\newblock {\em Theoretical population biology}, 38(2):219--232, 1990.

\bibitem{schelling1971dynamic}
Thomas~C Schelling.
\newblock Dynamic models of segregation.
\newblock {\em Journal of mathematical sociology}, 1(2):143--186, 1971.

\bibitem{axelrod1997complexity}
Robert~M Axelrod.
\newblock {\em The complexity of cooperation: Agent-based models of competition
  and collaboration}.
\newblock Princeton University Press, 1997.

\bibitem{reynolds1987flocks}
Craig~W Reynolds.
\newblock Flocks, herds and schools: A distributed behavioral model.
\newblock {\em ACM SIGGRAPH computer graphics}, 21(4):25--34, 1987.

\bibitem{pastor2015epidemic}
Romualdo Pastor-Satorras, Claudio Castellano, Piet Van~Mieghem, and Alessandro
  Vespignani.
\newblock Epidemic processes in complex networks.
\newblock {\em Reviews of modern physics}, 87(3):925, 2015.

\bibitem{centola2007complex}
Damon Centola and Michael Macy.
\newblock Complex contagions and the weakness of long ties1.
\newblock {\em American journal of Sociology}, 113(3):702--734, 2007.

\bibitem{moran1958random}
Patrick Alfred~Pierce Moran.
\newblock Random processes in genetics.
\newblock In {\em Proceedings of the Cambridge Philosophical Society},
  volume~54, page~60, 1958.

\bibitem{clifford1973model}
Peter Clifford and Aidan Sudbury.
\newblock A model for spatial conflict.
\newblock {\em Biometrika}, 60(3):581--588, 1973.

\bibitem{holley1975ergodic}
Richard~A Holley and Thomas~M Liggett.
\newblock Ergodic theorems for weakly interacting infinite systems and the
  voter model.
\newblock {\em The annals of probability}, pages 643--663, 1975.

\bibitem{suchecki2005voter}
Krzysztof Suchecki, V{\'\i}ctor~M Egu{\'\i}luz, and Maxi San~Miguel.
\newblock Voter model dynamics in complex networks: Role of dimensionality,
  disorder, and degree distribution.
\newblock {\em Physical Review E}, 72(3):036132, 2005.

\bibitem{castellano2005effect}
Claudio Castellano.
\newblock Effect of network topology on the ordering dynamics of voter models.
\newblock In {\em AIP Conference Proceedings}, volume 779, pages 114--120.
  American Institute of Physics, 2005.

\bibitem{sood2005voter}
Vishal Sood and Sidney Redner.
\newblock Voter model on heterogeneous graphs.
\newblock {\em Physical review letters}, 94(17):178701, 2005.

\bibitem{sood2008voter}
Vishal Sood, Tibor Antal, and Sidney Redner.
\newblock Voter models on heterogeneous networks.
\newblock {\em Physical Review E}, 77(4):041121, 2008.

\bibitem{barr2004establishing}
Dale~J Barr.
\newblock Establishing conventional communication systems: Is common knowledge
  necessary?
\newblock {\em Cognitive Science}, 28(6):937--962, 2004.

\bibitem{latane1981psychology}
Bibb Latane.
\newblock The psychology of social impact.
\newblock {\em American psychologist}, 36(4):343, 1981.

\bibitem{nowak1990private}
Andrzej Nowak, Jacek Szamrej, and Bibb Latan{\'e}.
\newblock From private attitude to public opinion: A dynamic theory of social
  impact.
\newblock {\em Psychological Review}, 97(3):362, 1990.

\bibitem{lewenstein1992statistical}
Maciej Lewenstein, Andrzej Nowak, and Bibb Latan{\'e}.
\newblock Statistical mechanics of social impact.
\newblock {\em Physical Review A}, 45(2):763, 1992.

\bibitem{holyst2000phase}
Janusz~A Ho{\l}yst, Krzysztof Kacperski, and Frank Schweitzer.
\newblock Phase transitions in social impact models of opinion formation.
\newblock {\em Physica A: Statistical Mechanics and its Applications},
  285(1):199--210, 2000.

\bibitem{schweitzer2000modelling}
Frank Schweitzer and JA~Ho{\l}yst.
\newblock Modelling collective opinion formation by means of active brownian
  particles.
\newblock {\em The European Physical Journal B-Condensed Matter and Complex
  Systems}, 15(4):723--732, 2000.

\bibitem{nowak2006evolutionary}
Martin~A Nowak.
\newblock {\em Evolutionary dynamics}.
\newblock Harvard University Press, 2006.

\bibitem{cox1989coalescing}
J~Theodore Cox.
\newblock Coalescing random walks and voter model consensus times on the torus
  in zd.
\newblock {\em The Annals of Probability}, pages 1333--1366, 1989.

\bibitem{baronchelli2007nonequilibrium}
Andrea Baronchelli, Luca DallÕAsta, Alain Barrat, and Vittorio Loreto.
\newblock Nonequilibrium phase transition in negotiation dynamics.
\newblock {\em Physical Review E}, 76(5):051102, 2007.

\bibitem{baronchelli2008depth}
Andrea Baronchelli, Vittorio Loreto, and Luc Steels.
\newblock In-depth analysis of the naming game dynamics: the homogeneous mixing
  case.
\newblock {\em International Journal of Modern Physics C}, 19(05):785--812,
  2008.

\bibitem{baronchelli2006topology}
Andrea Baronchelli, Luca DallÕAsta, Alain Barrat, and Vittorio Loreto.
\newblock Topology-induced coarsening in language games.
\newblock {\em Physical Review E}, 73(1):015102, 2006.

\bibitem{ellison1993learning}
Glenn Ellison.
\newblock Learning, local interaction, and coordination.
\newblock {\em Econometrica: Journal of the Econometric Society}, pages
  1047--1071, 1993.

\bibitem{travers1967small}
Jeffrey Travers and Stanley Milgram.
\newblock The small world problem.
\newblock {\em Phychology Today}, 1:61--67, 1967.

\bibitem{watts1998collective}
Duncan~J Watts and Steven~H Strogatz.
\newblock Collective dynamics of Ôsmall-worldÕnetworks.
\newblock {\em nature}, 393(6684):440--442, 1998.

\bibitem{barabasi1999emergence}
Albert-L{\'a}szl{\'o} Barab{\'a}si and R{\'e}ka Albert.
\newblock Emergence of scaling in random networks.
\newblock {\em Science}, 286(5439):509--512, 1999.

\bibitem{newman2003structure}
Mark~EJ Newman.
\newblock The structure and function of complex networks.
\newblock {\em SIAM review}, 45(2):167--256, 2003.

\bibitem{caldarelli2007scale}
Guido Caldarelli.
\newblock {\em Scale-free networks: complex webs in nature and technology}.
\newblock Oxford University Press, 2007.

\bibitem{dall2006nonequilibrium}
Luca DallÕAsta, Andrea Baronchelli, Alain Barrat, and Vittorio Loreto.
\newblock Nonequilibrium dynamics of language games on complex networks.
\newblock {\em Physical Review E}, 74(3):036105, 2006.

\bibitem{dall2006agreement}
Luca Dall'Asta, Andrea Baronchelli, Alain Barrat, and Vittorio Loreto.
\newblock Agreement dynamics on small-world networks.
\newblock {\em EPL (Europhysics Letters)}, 73:969, 2006.

\bibitem{vazquez2008analytical}
Federico Vazquez and V{\'\i}ctor~M Egu{\'\i}luz.
\newblock Analytical solution of the voter model on uncorrelated networks.
\newblock {\em New Journal of Physics}, 10(6):063011, 2008.

\bibitem{newmannetworks}
Mark Newman.
\newblock {\em Networks: an introduction.}
\newblock Oxford University Press Inc., New York, 2010.

\bibitem{montanari2010spread}
Andrea Montanari and Amin Saberi.
\newblock The spread of innovations in social networks.
\newblock {\em Proceedings of the National Academy of Sciences},
  107(47):20196--20201, 2010.

\bibitem{galam2008sociophysics}
Serge Galam.
\newblock Sociophysics: a review of galam models.
\newblock {\em International Journal of Modern Physics C}, 19(03):409--440,
  2008.

\bibitem{kuran1989sparks}
Timur Kuran.
\newblock Sparks and prairie fires: A theory of unanticipated political
  revolution.
\newblock {\em Public choice}, 61(1):41--74, 1989.

\bibitem{haken1975cooperative}
Hermann Haken.
\newblock Cooperative phenomena in systems far from thermal equilibrium and in
  nonphysical systems.
\newblock {\em Reviews of Modern Physics}, 47(1):67, 1975.

\bibitem{gladwell2006tipping}
Malcolm Gladwell.
\newblock {\em The tipping point: How little things can make a big difference}.
\newblock Little, Brown, 2006.

\bibitem{cavagna2016non}
Andrea Cavagna, Irene Giardina, Asja Jelic, Edmondo Silvestri, and Massimiliano
  Viale.
\newblock Non-symmetric interactions trigger collective swings in globally
  ordered systems.
\newblock {\em arXiv preprint arXiv:1605.00986}, 2016.

\bibitem{gladwell2010small}
Malcolm Gladwell.
\newblock Small change.
\newblock {\em The New Yorker}, 4(2010):42--49, 2010.

\bibitem{lighter1994random}
Jonathan~E Lighter and Random House.
\newblock {\em Random House historical dictionary of American slang}.
\newblock Random House, 1994.

\bibitem{galam2007role}
Serge Galam and Frans Jacobs.
\newblock The role of inflexible minorities in the breaking of democratic
  opinion dynamics.
\newblock {\em Physica A: Statistical Mechanics and its Applications},
  381:366--376, 2007.

\bibitem{mobilia2003does}
Mauro Mobilia.
\newblock Does a single zealot affect an infinite group of voters?
\newblock {\em Physical review letters}, 91(2):028701, 2003.

\bibitem{mobilia2003majority}
M~Mobilia and S~Redner.
\newblock Majority versus minority dynamics: Phase transition in an interacting
  two-state spin system.
\newblock {\em Physical Review E}, 68(4):046106, 2003.

\bibitem{xie2011social}
Jierui Xie, Sameet Sreenivasan, Gyorgy Korniss, Weituo Zhang, Chjan Lim, and
  Boleslaw~K Szymanski.
\newblock Social consensus through the influence of committed minorities.
\newblock {\em Physical Review E}, 84(1):011130, 2011.

\bibitem{marvel2012encouraging}
Seth~A Marvel, Hyunsuk Hong, Anna Papush, and Steven~H Strogatz.
\newblock Encouraging moderation: clues from a simple model of ideological
  conflict.
\newblock {\em Physical review letters}, 109(11):118702, 2012.

\bibitem{mistry2015committed}
Dina Mistry, Qian Zhang, Nicola Perra, and Andrea Baronchelli.
\newblock Committed activists and the reshaping of status-quo social consensus.
\newblock {\em Physical Review E}, 92(4):042805, 2015.

\bibitem{halu2013connect}
Arda Halu, Kun Zhao, Andrea Baronchelli, and Ginestra Bianconi.
\newblock Connect and win: The role of social networks in political elections.
\newblock {\em EPL (Europhysics Letters)}, 102(1):16002, 2013.

\bibitem{niu2017impact}
Xiang Niu, Casey Doyle, Gyorgy Korniss, and Boleslaw~K Szymanski.
\newblock The impact of variable commitment in the naming game on consensus
  formation.
\newblock {\em Scientific Reports}, 7:41750, 2017.

\bibitem{shao2009dynamic}
Jia Shao, Shlomo Havlin, and H~Eugene Stanley.
\newblock Dynamic opinion model and invasion percolation.
\newblock {\em Physical review letters}, 103(1):018701, 2009.

\bibitem{mas2010individualization}
Michael M{\"a}s, Andreas Flache, and Dirk Helbing.
\newblock Individualization as driving force of clustering phenomena in humans.
\newblock {\em PLoS Comput Biol}, 6(10):e1000959, 2010.

\bibitem{abelson1964mathematical}
Robert~P Abelson.
\newblock Mathematical models of the distribution of attitudes under
  controversy.
\newblock {\em Contributions to mathematical psychology}, 14:1--160, 1964.

\bibitem{li2013non}
Qian Li, Lidia~A Braunstein, Huijuan Wang, Jia Shao, H~Eugene Stanley, and
  Shlomo Havlin.
\newblock Non-consensus opinion models on complex networks.
\newblock {\em Journal of Statistical Physics}, 151(1-2):92--112, 2013.

\bibitem{lambiotte2007coexistence}
R~Lambiotte and Marcel Ausloos.
\newblock Coexistence of opposite opinions in a network with communities.
\newblock {\em Journal of Statistical Mechanics: Theory and Experiment},
  2007(08):P08026, 2007.

\bibitem{lu2009naming}
Qiming Lu, Gyorgy Korniss, and Boleslaw~K Szymanski.
\newblock The naming game in social networks: community formation and consensus
  engineering.
\newblock {\em Journal of Economic Interaction and Coordination}, 4(2):221,
  2009.

\bibitem{schweitzer2009nonlinear}
Frank Schweitzer and Laxmidhar Behera.
\newblock Nonlinear voter models: the transition from invasion to coexistence.
\newblock {\em The European Physical Journal B-Condensed Matter and Complex
  Systems}, 67(3):301--318, 2009.

\bibitem{axelrod1997dissemination}
Robert Axelrod.
\newblock The dissemination of culture a model with local convergence and
  global polarization.
\newblock {\em Journal of conflict resolution}, 41(2):203--226, 1997.

\bibitem{deffuant2000mixing}
Guillaume Deffuant, David Neau, Frederic Amblard, and G{\'e}rard Weisbuch.
\newblock Mixing beliefs among interacting agents.
\newblock {\em Advances in Complex Systems}, 3(01n04):87--98, 2000.

\bibitem{castellano2000nonequilibrium}
Claudio Castellano, Matteo Marsili, and Alessandro Vespignani.
\newblock Nonequilibrium phase transition in a model for social influence.
\newblock {\em Physical Review Letters}, 85(16):3536, 2000.

\bibitem{klemm2003nonequilibrium}
Konstantin Klemm, V{\'\i}ctor~M Egu{\'\i}luz, Ra{\'u}l Toral, and Maxi
  San~Miguel.
\newblock Nonequilibrium transitions in complex networks: A model of social
  interaction.
\newblock {\em Physical Review E}, 67(2):026120, 2003.

\bibitem{gonzalez2005nonequilibrium}
Juan~Carlos Gonz{\'a}lez-Avella, Mario~G Cosenza, and Kay Tucci.
\newblock Nonequilibrium transition induced by mass media in a model for social
  influence.
\newblock {\em Physical Review E}, 72(6):065102, 2005.

\bibitem{gonzalez2006local}
Juan~Carlos Gonz{\'a}lez-Avella, V{\'\i}ctor~M Egu{\'\i}luz, Mario~G Cosenza,
  Konstantin Klemm, Jose~L Herrera, and Maxi San~Miguel.
\newblock Local versus global interactions in nonequilibrium transitions: A
  model of social dynamics.
\newblock {\em Physical Review E}, 73(4):046119, 2006.

\bibitem{centola2007homophily}
Damon Centola, Juan~Carlos Gonzalez-Avella, Victor~M Eguiluz, and Maxi
  San~Miguel.
\newblock Homophily, cultural drift, and the co-evolution of cultural groups.
\newblock {\em Journal of Conflict Resolution}, 51(6):905--929, 2007.

\bibitem{del2016echo}
Michela Del~Vicario, Gianna Vivaldo, Alessandro Bessi, Fabiana Zollo, Antonio
  Scala, Guido Caldarelli, and Walter Quattrociocchi.
\newblock Echo chambers: Emotional contagion and group polarization on
  facebook.
\newblock {\em Scientific Reports}, 6, 2016.

\bibitem{del2016spreading}
Michela Del~Vicario, Alessandro Bessi, Fabiana Zollo, Fabio Petroni, Antonio
  Scala, Guido Caldarelli, H~Eugene Stanley, and Walter Quattrociocchi.
\newblock The spreading of misinformation online.
\newblock {\em Proceedings of the National Academy of Sciences},
  113(3):554--559, 2016.

\bibitem{starnini2016emergence}
Michele Starnini, Mattia Frasca, and Andrea Baronchelli.
\newblock Emergence of metapopulations and echo chambers in mobile agents.
\newblock {\em Scientific reports}, 6, 2016.

\bibitem{howell2017global}
Lee Howell.
\newblock Global risks report 2017.
\newblock In {\em World Economic Forum}, 2017.

\bibitem{zollo2015debunking}
Fabiana Zollo, Alessandro Bessi, Michela Del~Vicario, Antonio Scala, Guido
  Caldarelli, Louis Shekhtman, Shlomo Havlin, and Walter Quattrociocchi.
\newblock Debunking in a world of tribes.
\newblock {\em arXiv preprint arXiv:1510.04267}, 2015.

\bibitem{galantucci2005experimental}
Bruno Galantucci.
\newblock An experimental study of the emergence of human communication
  systems.
\newblock {\em Cognitive science}, 29(5):737--767, 2005.

\bibitem{selten2007emergence}
Reinhard Selten and Massimo Warglien.
\newblock The emergence of simple languages in an experimental coordination
  game.
\newblock {\em Proceedings of the National Academy of Sciences},
  104(18):7361--7366, 2007.

\bibitem{lewis1975languages}
David Lewis.
\newblock Languages and language.
\newblock {\em Minnesota studies in the philosophy of science}, 7:3--35, 1975.

\bibitem{garrod1987saying}
Simon Garrod and Anthony Anderson.
\newblock Saying what you mean in dialogue: A study in conceptual and semantic
  co-ordination.
\newblock {\em Cognition}, 27(2):181--218, 1987.

\bibitem{centola2015spontaneous}
Damon Centola and Andrea Baronchelli.
\newblock The spontaneous emergence of conventions: An experimental study of
  cultural evolution.
\newblock {\em Proceedings of the National Academy of Sciences},
  112(7):1989--1994, 2015.

\bibitem{judd2010behavioral}
Stephen Judd, Michael Kearns, and Yevgeniy Vorobeychik.
\newblock Behavioral dynamics and influence in networked coloring and
  consensus.
\newblock {\em Proceedings of the National Academy of Sciences},
  107(34):14978--14982, 2010.

\bibitem{centola2010spread}
Damon Centola.
\newblock The spread of behavior in an online social network experiment.
\newblock {\em science}, 329(5996):1194--1197, 2010.

\bibitem{christakis2007spread}
Nicholas~A Christakis and James~H Fowler.
\newblock The spread of obesity in a large social network over 32 years.
\newblock {\em New England journal of medicine}, 357(4):370--379, 2007.

\bibitem{hodas2013simple}
Nathan~O Hodas and Kristina Lerman.
\newblock The simple rules of social contagion.
\newblock {\em Scientific Reports}, 4.

\bibitem{aral2017}
Sinan Aral and Christos Nicolaides.
\newblock Exercise contagion in a global social network.
\newblock {\em Nature Communications}, 8.

\bibitem{hawkins2016formation}
Robert~XD Hawkins and Robert~L Goldstone.
\newblock The formation of social conventions in real-time environments.
\newblock {\em PloS one}, 11(3):e0151670, 2016.

\bibitem{kooti2012emergence}
Farshad Kooti, Haeryun Yang, Meeyoung Cha, P~Krishna Gummadi, and Winter~A
  Mason.
\newblock The emergence of conventions in online social networks.
\newblock In {\em ICWSM}, 2012.

\bibitem{heaberlin2016evolution}
Bradi Heaberlin and Simon DeDeo.
\newblock The evolution of wikipediaÕs norm network.
\newblock {\em Future Internet}, 8(2):14, 2016.

\bibitem{haken}
H~Haken.
\newblock {\em Information and Self-Organisation: a macroscopic approach to
  complex systems}.
\newblock Springer-Verlag, Berlin, 1988.

\bibitem{albert2002statistical}
R{\'e}ka Albert and Albert-L{\'a}szl{\'o} Barab{\'a}si.
\newblock Statistical mechanics of complex networks.
\newblock {\em Reviews of modern physics}, 74(1):47, 2002.

\bibitem{boccaletti2006complex}
Stefano Boccaletti, Vito Latora, Yamir Moreno, Martin Chavez, and D-U Hwang.
\newblock Complex networks: Structure and dynamics.
\newblock {\em Physics reports}, 424(4):175--308, 2006.

\bibitem{fortunato2010community}
Santo Fortunato.
\newblock Community detection in graphs.
\newblock {\em Physics reports}, 486(3):75--174, 2010.

\end{thebibliography}
\end{document}